\documentclass[adp,fleqn]{w-art}
\usepackage{times}
\usepackage{w-thm}
\usepackage{epsfig}
\usepackage{amsmath}
\usepackage{sidecap}

\begin{document}
%\authorrunning{F. Fehrer et al} 
\title{Modeling Na clusters in Ar matrices}
\author{F. Fehrer\inst{1}}
\address[\inst{1}]
{Institut f{\"u}r Theoretische Physik, Universit{\"a}t Erlangen,
   Staudtstrasse 7, D-91058 Erlangen,
   Germany}
\author{M. Mundt\inst{1,2}}
\address[\inst{2}]
{Laboratoire de Physique Th\'eorique,
   Universit{\'e} Paul Sabatier,
  118 Route de Narbonne, F-31062 Toulouse, cedex, France }
\author{P.-G. Reinhard\inst{1,2}}
\author{E. Suraud\inst{2}}
%\date{1. Draft: 8. August 2004} 
%
%\abstract{ 
\begin{abstract}
We present a microscopic model for Na clusters embedded in raregas
matrices. The valence electrons of the Na cluster are described by
time-dependent density-functional theory at the level of the local-density
approximation (LDA). Particular attention is paid to the semi-classical 
picture in terms of Vlasov-LDA. The Na ions and Ar atoms are handled
as classical particles whereby the Ar atoms carry two degrees of freedom,
position and dipole polarization. The interaction between Na ions and
electrons is mediated through local pseudo-potentials. The coupling to the Ar
atoms is described by (long-range) polarization potentials and (short-range)
repulsive cores. The ingredients are taken from elsewhere developed
standards. A final fine-tuning is performed using the NaAr molecule as
benchmark. 
The model is then applied to embedded systems  Na$_8$\@Ar$_N$.
By close comparison with quantum-mechanical results, we explore the
capability of the Vlasov-LDA to describe such embedded clusters.
We show that one can obtain a reasonable description
by appropriate adjustments in the fine-tuning phase of the model. 
\end{abstract}
\maketitle 
%} 
%
%\PACS{??}  

\section{Introduction}

Structure and dynamics of metal clusters has been a much studied topic over the
last two decades, see e.g. \cite{Kre93,Hab94a,Eka99}. While optical response
was in the focus of numerous  earlier studies, recent developments aim at exploring
cluster dynamics in all its aspects \cite{Rei03a}. The preferred theoretical
tool for describing clusters in all dynamical regimes is the time-dependent
local-density approximation (TDLDA) which treats the cluster electrons at a
fully quantum mechanical level, however in an independent-particle picture as
motion in a common self-consistent mean-field. 
For highly excited systems, one can employ a
semi-classical approximation in terms of phase-space dynamics described by the
Vlasov equation, in the cluster context called a Vlasov-LDA. The Vlasov
equation was originally used in a genuinely classical environment for plasma
physics \cite{Vla50}. Nuclear physicists have adapted it to finite quantum
systems for the description of violent heavy-ion collisions \cite{Ber88}.  The
motivation is twofold: first, the Vlasov description becomes more efficient
for highly excited system, and second, the semi-classical approach makes it
feasible to include correlations simply in terms of an \"Uhling-Uhlenbeck collision
term (extended Vlasov to VUU). The successful applications in nuclear dynamics
have motivated the transfer of the method to cluster physics, in the first
round using a jellium model for the ionic background \cite{Fer96,Pla99,Dom00a}
but later also in connection with detailed ions \cite{Gig01a,Fen04}. These
applications of Vlasov-LDA and VUU dealt up to now only with free metal
clusters. Many experiments, however, are done on supported or embedded
clusters, see for example the study of the systematics of optical properties
in small Ag clusters \cite{Die02}. There is thus a need to develop reliable
and efficient models for the cluster-substrate interface in connection with
cluster dynamics in various regimes. Such developments are underway for TDLDA
in embedded clusters whereby raregas matrices are considered as a substrate
with simple constituents \cite{xx1,xx2,Feh04a}. They take up the modeling of
the Ar atoms and their interaction with the Na cluster from earlier
quantum-chemical considerations \cite{Dup96}. It is the aim of the present
paper to investigate the performance of Vlasov-LDA and VUU for metal clusters
embedded in raregas matrices. To that end, we recycle the interface model from
the TDLDA studies and combine it with the Vlasov-LDA description for the
cluster electrons. We intend to develop a parameterization of the model which
provides acceptable ground state structures. These are then used as laboratory
for dynamical studies. In the first exploration here, we use a simple
excitation mechanism (instantaneous boost) to evaluate the basic features of
dynamical response in various regimes, namely spectral distributions and
ionization. The validity and limitations of Vlasov-LDA are scrutinized by
comparison with results from quantum mechanical TDLDA. We also exploit the
option of VUU for a first exploration of dynamical correlations and their
interlay with the substrate.

The paper is outlined as follows: 
In section \ref{sec:enfundetail}, we present the model in detail thereby
providing all information for full TDLDA as well as for the semi-classical
approaches.
In section \ref{sec:solve}, we summarize briefly the emerging equations of
motion and their numerical solution.
In section \ref{sec:benchNaAr}, we discuss the calibration of the model using
the NaAr molecule as benchmark.
Finally in section \ref{sec:Na8}, we present and discuss first results for an
embedded cluster with Na$_8$\@Ar$_N$ as test cases.

\section{The Na-Ar energy functional in detail}
\label{sec:enfundetail}

\subsection{The degrees of freedom}

The key constituents are the valence electrons of the metal cluster.  Starting
point for their description is quantum mechanical mean-field theory where each
electron is associated with one single-particle wavefunction $\{\varphi_n({\bf
r}),n=1...N_{\rm el}\}$.  The $\varphi_n$ are determined by a (time-dependent)
Kohn-Sham equation.  The manifold of occupied wavefunctions is comprised in
the one-electron density matrix
$\rho({\bf r},{\bf r}')=
 \sum_n\varphi_n^{\mbox{}}({\bf r})\varphi_n^+({\bf r}')$.
We perform a semi-classical approximation in order to obtain a description in
terms of the one-electron phase-space distribution $f({\bf r},{\bf p})$
\cite{Dom97b}. This yields the Vlasov-LDA which has shown to provide a
reliable description of dynamical processes in free metal clusters with
detailed ionic structure \cite{Gig02,Gig03}.
The other ingredients are the Na ions in the cluster and the raregas
atoms. They are treated by classical molecular dynamics in terms of positions
${\bf R}$ and associated momenta ${\bf P}$. The Ar atom carries as internal
degree of freedom a dipole moment to account for the polarization
interaction. Effective Ar potentials for static calculations contain that
implicitly. True dynamics requires explicit dipoles as we use it here.  It is
convenient to formulate the Ar dipoles ${\bf d}_a$ in terms of the
displacement
${\bf R}'_a-{\bf R}^{\mbox{}}_a = {\bf d}_a/e q_{\rm Ar}$.  
All these degrees of freedom are summarized in table \ref{tab:dof}.
\begin{table}
\begin{center}
\begin{tabular}{|lcl|}
\hline
 electrons    &$\longleftrightarrow$
              & $\{\varphi_n({\bf r}),n\!=\!1...N_{\rm el}\}
                \quad\Longrightarrow\quad f({\bf r},{\bf p})$\\
 cluster ions &$\longleftrightarrow$
              & $\{{\bf R}_I,I=1...N_{\rm ion}\}$ \\
\hline
 Ar core   &$\longleftrightarrow$
              & $\{{\bf R}_a,a=1...N_{\rm Ar}\}$ \\
 Ar valence cloud&$\longleftrightarrow$
              & ${\bf R'}_a={\bf R}_a+{\bf s}_a$ \\
\hline
\end{tabular}
\end{center}
\caption{\label{tab:dof}
Short summary of the degrees of freedom of the model.
Note that the center of the valence cloud ${\bf R'}_a$
serves to characterize the Ar dipole moment
through ${\bf d}=eq_{\rm Ar}{\bf s}=eq_{\rm Ar}({\bf R}'-{\bf R})$ 
where $q_{\rm Ar}$ is the net charge of the valence cloud.
}
\end{table}

The detailed description of the Ar atoms is outlined in the next subsection
\ref{sec:Ardetails}.  What the Na$^+$ ions is concerned, we neglect the small
dipole polarizability of the Na$^+$ core and treat it merely as a
monopole. The ions are taken as point charges in the interaction with the Ar
atoms. Soft and local pseudo-potentials are used for the interaction with the
cluster electrons \cite{Kue99}, see eq. (\ref{eq:PsP}).
The density of the electrons is given naturally as defined in
mean-field theories. It reads in the quantum mechanical version
\begin{subequations}
\begin{equation}
  \rho_{\rm el}({\bf r})
  =
  \sum_\alpha\left|\varphi_\alpha^{\mbox{}}({\bf r})\right|^2
\end{equation}
and in Vlasov-LDA
\begin{equation}
  \rho_{\rm el}({\bf r})
  =
  \int d^3p\,f({\bf r},{\bf p})
  \quad.
\label{eq:rhoelV}
\end{equation}
\end{subequations}

\subsection{Construction of soft dipole potentials}
\label{sec:Ardetails}

A finite dipole momentum in Ar atom is generated from Coulomb fields of the
other constituents or from the excitation mechanism.  In turn, it produces a
dipole field to the outside world. However, a pure dipole field has a huge
singularity at the origin. This is unphysical because we are not having point
dipoles in practice. With a closer look at the Ar atom we see that the dipole
is generated from a deformation (more precisely, displacement) of the outer
electron shell ($3s$-$3p$-shell) and the finite size of the source delivers
regular Coulomb fields everywhere. We thus specify the dipole momentum
of the Ar atom in more detail.
The Ar atom is thought of consisting out of a valence electron
cloud with center of mass ${\bf R}'_a$ and of the corresponding
positively charged core placed at the atom position ${\bf
R}_a^{\mbox{}}$.  (In fact, one should use the center-of-mass of core
plus valence cloud but the recoil effect on the Ar core is negligible,
thus neglected). Both, valence cloud and core have the same absolute
value of charge $q_{\rm Ar}$ delivering together a neutral atom. A
small displacement of the valence electrons against the core produces
a net dipole moment. The profile of valence cloud and core is
described by the same Gaussian with width $\sigma_{\rm Ar}$, yielding
together the charge distribution of one Ar atom as
\begin{subequations}
\begin{equation} 
  \rho_{{\rm Ar},a}({\bf r})
  =
  \frac{e q_{\rm Ar}}{\pi^{3/2}_{\mbox{}}\sigma_{\rm Ar}^3}
  \Big[
   \exp{\left(-\frac{({\bf r}\!-\!{\bf R}^{\mbox{}}_a)^2}{\sigma_{\rm Ar}^2}\right)}
   -
   \exp{\left(-\frac{({\bf r}\!-\!{\bf R}'_a)^2}
                    {\sigma_{\rm Ar}^2}\right)}
  \Big]
\label{eq:Ardistri}
\end{equation}
where the index $a$ stands for Ar atom number $a$ implying atom
position ${\bf R}^{\mbox{}}_a$ and valence cloud ${\bf R}'_a$.
The Coulomb potential of this charge density produces the polarization-part
of the potential exerted from the Ar atoms,
\begin{equation} 
  V^{\rm(pol)}_{{\rm Ar},a}({\bf r})
  =
  e^2{q_{\rm Ar}^{\mbox{}}}
  \Big[
   \frac{\mbox{erf}\left(|{\bf r}\!-\!{\bf R}^{\mbox{}}_a|
          /\sigma_{\rm Ar}^{\mbox{}}\right)}
        {|{\bf r}\!-\!{\bf R}^{\mbox{}}_a|}
   -
   \frac{\mbox{erf}\left(|{\bf r}\!-\!{\bf R}'_a|/\sigma_{\rm Ar}^{\mbox{}}\right)}
        {|{\bf r}\!-\!{\bf R}'_a|}
  \Big]
  \quad,
\label{eq:Arpolpot}
\end{equation}
where $q_{\rm Ar}^{\mbox{}}$ is the effective charge of the 
Ar cloud and core (see below). The error function therein is defined as
\begin{equation} 
  \mbox{erf}(r)
  = 
  \frac{2}{\sqrt{\pi}}\int_0^r dx\,e^{-x^2}
  \quad.
\label{eq:erf}
\end{equation}
\end{subequations}

\subsection{The total energy}

The model is fully specified by giving the total energy of the system.
It is composed as
%\widetext
\begin{subequations}
\label{eq:etotal}
\begin{equation}
  E_{\rm total}
  =
  \underbrace{E_{\rm Na,el}
  +
  E_{\rm Na,ion}
  +
  E_{\rm ion,el}}_{\mbox{\tiny Na cluster}}
  +
  E_{\rm Ar}
  +
  E_{\rm coupl}
  +
  E_{\rm VdW}
  \quad,
\end{equation}
\begin{eqnarray}
  E_{\rm Na,el}
  &=&
  E_{\rm kin}
  +
  E_{\rm Hartree}(\rho_{\rm el})
%  \frac{e^2}{2} \int{d{\bf r}\,{d{\bf r}' 
%  \frac{\rho_{\rm el}({\bf r})\rho_{\rm el}({\bf r}')}
%       {\left|{\bf r}-{\bf r}'\right|}}}
  +
  E_{\rm xc}(\rho_{\rm el})
  +
  \int{d{\bf r} V_{\rm ext}({\bf r},t) \rho_{\rm el}({\bf r})}
  \quad,
\\
  E_{\rm Na,ion}
  &=&  
  \sum_I \frac{{\bf P}_I^2}{2M_{\rm Na}}
  +
  \sum_{I<J} \frac{e^2}{|{\bf R}_I - {\bf R}_J|}
  \quad,
\\
  E_{\rm ion,el}
  &=&
  \sum_I\int{d{\bf r} V_{\rm PsP}(|{\bf r}-{\bf R}_I|)
                      \rho_{\rm el}({\bf r})}
  \quad,
\label{eq:Naelcoup}\\
  E_{\rm Ar}
  &=&
  \sum_a \frac{{\bf P}_a^2}{2M_{\rm Ar}} 
  +
  \sum_a \frac{{{\bf P}'_{a}}^2}{2m_{\rm Ar}}
  +
  \frac{1}{2} k_{\rm Ar}\left({\bf R}'_{a}-{\bf R}_{a}\right)^2
\nonumber\\
  &&  
  +
  \sum_{a<a'}
  \left[
    \int d{\bf r}\rho_{{\rm Ar},a}({\bf r})
    V^{\rm(pol)}_{{\rm Ar},a'}({\bf r})
    +
    V^{\rm(core)}_{\rm ArAr}({\bf R}_a - {\bf R}_{a'})
  \right]
\nonumber\\
  &&  
  +
  \sum_a \int d{\bf r} V_{\rm ext}({\bf r},t) 
  \rho_{{\rm Ar},a}({\bf r})
  \quad,
\\
  E_{\rm coupl}
  &=&
  \sum_{I,a}\left[
    V^{\rm(pol)}_{{\rm Ar},a}({\bf R}_{I})
    +
    V'_{\rm NaAr}({\bf R}_I - {\bf R}_a)
  \right]
\nonumber\\
  &&
  +
  \int d{\bf r}\rho_{\rm el}({\bf r})\sum_a \left[
    V^{\rm(pol)}_{{\rm Ar},a}({\bf r})
    +
    W_{\rm elAr}(|{\bf r}-{\bf R}_a|)
  \right]
  \quad,
\\
  E_{\rm VdW}
  &=&  
  e^2\frac{1}{2} \sum_a \alpha_a
  \Big[\frac{1}{N_{\rm el}}
    \left(\int{d{\bf r} {\bf f}_a({\bf r}) \rho_{\rm el}({\bf r})}\right)^2
      - \int{d{\bf r} {\bf f}_a({\bf r})^2 \rho_{\rm el}({\bf r})}
  \Big]
  \;,
\label{eq:EvdW}
\\
  {\bf f}_a({\bf r})
  &=&
  \nabla\frac{\mbox{erf}\left(|{\bf r}\!-\!{\bf R}^{\mbox{}}_a|
          /\sigma_{\rm Ar}^{\mbox{}}\right)}
        {|{\bf r}\!-\!{\bf R}^{\mbox{}}_a|}
  \quad.
\label{eq:effdip}
\end{eqnarray}
The electronic kinetic energy depends on the level of approach, i.e.
\begin{equation}
  E_{\rm kin}
  =
  \sum_n\left|\nabla\varphi_n\right|^2
\end{equation}
in  the quantum case and 
\begin{equation}
  E_{\rm kin}
  =
  \int d^3p\,\frac{{\bf p}^2}{2m_{\rm el}}f({\bf r},{\bf p})
  \quad.  
\end{equation}
\end{subequations}
in the Vlasov case.
The Van-der-Waals energy $E_{\rm VdW}$ is a correlation from the dipole
excitation in the Ar atom coupled with a dipole excitation in the cluster. It
requires, in principle, an energy denominator involving the basic excitation
energy $\Delta E_{\rm Ar}$ in the Ar atom and the typical excitation energy in
the Na cluster which is $\approx\omega_{\rm Mie}$. We exploit the fact that
$\omega_{\rm Mie}\ll\Delta E_{\rm Ar}$ and eliminate the energy
denominator in the spirit of a closure approximation. This amounts to
compute the variance of the dipole operator in the cluster. It employs, in
fact, an effective dipole operator ${\bf f}_a$, see eq.  (\ref{eq:effdip}),
which corresponds to the dipole field from the smoothened Ar charge
distributions.
The dipole variance as expectation value within the many-body wavefunctions
is, furthermore, simplified in terms of the local variance.  This yields
finally the approximation (\ref{eq:EvdW}) to the Van-der-Waals energy. Its
computation is still very cumbersome. It becomes particularly involved in the
semi-classical case. On the other hand, Vlasov-LDA is intended for the regime
of high excitations where subtle details, as e.g. the Van-der-Waals force
do not matter. Thus we skip $E_{\rm VdW}$ in Vlasov-LDA.

It is obvious that the polarization potentials are handled in terms of
effective smoothened pseudo-densities (\ref{eq:Ardistri}) and their resulting
Coulomb potential (\ref{eq:Arpolpot}). The repulsive core potentials require
explicit modeling. We take it mainly from literature.  The contributions
are
\begin{subequations}
\begin{eqnarray}
  W_{\rm elAr}(r)
  &=&
  e^2\frac{A_{\rm el}}{1+e^{\beta_{\rm el}(r - r_{\rm el})}}
%  e^2\frac{2.3}{1+e^{2.8(r_b - 1.8)}}
\label{eq:VArel}\\
  V_{\rm ArAr}^{\rm (core)}(R)
  &=& 
  e^2 \, 1.367 \cdot 10^{-3}\Bigg[
  \left( \frac{6.501}{R}\right)^{12}
 -\left( \frac{6.501}{R}\right)^{6}
  \!\Bigg]
\label{eq:VArAr}
\\
  V'_{\rm ArNa}(R)
  &=&
  e^2\Bigg[
  A_{\rm Na} \frac{e^{-\beta_{\rm Na} R}}{R}
  -
  \frac{2}{1+e^{\alpha_{\rm Na}/R}}
  \left(\frac{C_{\rm Na,6}}{R^6} + \frac{C_{\rm Na,8}}{R^8}\right)
  \Bigg]
\label{eq:VpArNa}
\end{eqnarray}
\end{subequations}
\begin{table}[h!]
\begin{center}
\begin{tabular}{|l|ccccc|}
\hline
  & $\beta_{\rm Na}$ & $\alpha_{\rm Na}$ & $A_{\rm Na}$ & $C_{\rm{Na},6}$ & $C_{\rm Na,8}$\\
\hline
 quantum mechanical LDA & 1.7624 a$_0^{-1}$ & 1.815 a$_0$ & 334.85 Ha & 52.5 Ha\;a$_0^6$ & 1383 Ha\;a$_0^8$ \\ 
 Vlasov-LDA & 1.670 a$_0^{-1}$ & 5.6 a$_0$ & 242.32 Ha & 225 Ha\;a$_0^6$ & 4410 Ha\; a$_0^8$\\
\hline
\end{tabular}
\end{center}
\caption{\label{tab:ArNapot}
Parameters for the ion-Ar pseudo-potential in Na
according to the form (\ref{eq:VpArNa}).
}
\end{table}

For the Ar-Ar core interaction we employ a Lennard-Jones type
potential with parameters such that binding properties of bulk Ar are
reproduced.  The Na-Ar core potential has been fitted according to
\cite{Rez95}. Note that the Na-Ar potential from \cite{Rez95} does
contain also a long range part $\propto\alpha_{\rm Ar}$ which accounts
for the dipole polarization-potential. We describe that long range part
explicitely and have to omit it here. We thus choose the form 
(\ref{eq:VpArNa}).
The pseudo-potential $W_{\rm elAr}$ for the electron-Ar core repulsion
has been modeled according to the proposal of \cite{Dup96}. Its
parameters will be re-tuned as discussed below.

\subsection{The ion-electron pseudo-potential}

The pseudo-potential for the coupling (\ref{eq:Naelcoup}) between Na ions and
electrons is taken in a soft and local form as \cite{Kue99}
\begin{equation}
  V_{\rm PsP}(r)
  =
  -\frac{e^2}{r}\bigg[
    c_1 \mbox{erf}\bigg(\frac{r}{\sqrt{2}\sigma_1}\bigg)
    +
    c_2 \mbox{erf}\bigg(\frac{r}{\sqrt{2}\sigma_2}\bigg)
   \bigg]
\label{eq:PsP}
\end{equation}
with the error function given in eq. (\ref{eq:erf}).  The parameters were
adjusted to the properties of Na atom and bulk.  This serves to provide
simultaneously correct bond lengths for Na molecule and clusters as well as as
good description of optical response, all in the framework of
quantum-mechanical (TD)LDA \cite{Kue99}. A semi-classical description of Na
clusters does also work very well with a pseudo-potential in the form
(\ref{eq:PsP}). But it requires a new adjustment of the parameters suited for
the Vlasov approach \cite{Gig02b}. Both version for the parameters are
summarized in table \ref{tab:PsP}.
\begin{table}
\begin{center}
\begin{tabular}{|l|cccc|}
\hline
  & $c_1$ & $c_2$ & $\sigma_1$ & $\sigma_2$\\
\hline
 quantum mechanical LDA & -2.29151 & 3.29151 & 0.681 a$_0$ & 1.163 a$_0$\\ 
 Vlasov-LDA & -0.5 & 1.5 & 0.7778 a$_0$ & 1.5556 a$_0$ \\
\hline
\end{tabular}
\end{center}
\caption{\label{tab:PsP}
Parameters for the ion-electron pseudo-potential in Na
according to the local form (\ref{eq:PsP}). Different sets are to be used for
quantum mechanical LDA and for semi-classical Vlasov-LDA.
}
\end{table}

\subsection{The parameters for the polarization potential}

The polarizability of the Ar atoms is described by the model
of a rigid cloud of valence electrons which is oscillating
against the remaining raregas ion.
The parameters of the model are:\\
\hspace*{2em}
\begin{tabular}{lcl}
$q_{\rm Ar}$      
  &=& effective charge of valence cloud\\
$m_{\rm Ar}$      &=& effective mass of valence cloud  \\
$k_{\rm Ar}$      &=& restoring force for dipoles\\
\end{tabular}
\\ 
These parameters are adjusted to reproduce the basic response
properties of the raregas atom in the low energy domain, i.e. we choose to
reproduce the static polarizability $\alpha_{RG}$ and the second
derivative of the dynamical polarizability
$\partial_\omega^2\alpha_D|_{\omega\!=\!0}$. To that end, we make a 
one-pole model having the dispersion relation
\begin{subequations}
\label{eq:dipparams}
\begin{equation}
  \alpha_{\rm Ar}(\omega)
  =
  \frac{\alpha_{\rm Ar}(0)}{1-\frac{\omega^2}{\omega_0^2}}
  \approx
  {\alpha_{\rm Ar}}+\frac{{\alpha_{\rm Ar}}}{\omega_0^2}\omega^2
  \quad.
\end{equation}
The frequency, in turn, is related to spring constant and mass 
as usual 
$$
  \omega_0=\sqrt{k_{\rm Ar}/m_{\rm Ar}}
  \quad.
$$
On the other hand, the spring constant is directly related to the static
polarizability as
\begin{equation}
  k_{\rm Ar}
  =
  \frac{e^2q_{\rm Ar}^2}{\alpha_{\rm Ar}}
  \quad.
\end{equation}
We identify the mass of the electron cloud with its charge and
the electron mass as
\begin{equation}
  m_{\rm Ar}
  =
  q_{\rm Ar}m_{\rm el}
  \quad.
\label{eq:dipmass}
\end{equation}
thus we find for the effective charge of the valence cloud
\begin{equation}
  q_{\rm Ar}
  =
  \frac{\alpha_{\rm Ar}m_{\rm el}\omega_0^2}{e^2}
  \quad.
\end{equation}
The folding width of the valence cloud is determined such that the
restoring force from the folded Coulomb (for small displacements)
reproduces the spring constant $k_{\rm Ar}$. This yields
\begin{equation}
  \sigma_{\rm RG}
  =
  \left(\alpha_{\rm Ar}\frac{4\pi}{3(2\pi)^{3/2}}  \right)^{1/3}
  \quad.
\end{equation}
For Ar we use 
\begin{equation}
  \fbox{$
  \alpha_{\rm Ar}(0) = 11.08\,{\rm a}_0^3
  \quad,\quad
  \omega_0 = 1.755\,{\rm Ry}
  \quad.
  $}
\end{equation}
\end{subequations}
The eqs. (\ref{eq:dipparams}) determine the parameters 
for the polarization potentials at the side of the Ar atoms.

\subsection{Fine tuning of electron-Ar core potential}
\label{sec:tuneArel}

The parameters of the electron-Ar core potential (\ref{eq:VArel}) determine
sensitively the binding properties of Na to the Ar atoms.  We use that
potential as a means for a final fine-tuning of the model. The benchmark for
adjustment is provided by the Na-Ar dimer. The data are taken from
\cite{Gro98} and \cite{Rho02a} and summarized in table \ref{tab:NaArdata}.
\begin{SCtable}[0.5]
%\begin{center}
\begin{tabular}{| l| l| }
\hline 
binding distance $d_0$ & 9.5 $a_0$  \\
depth $E_0$ of the $X^2 \Sigma^+$ ground-state & ${\rm 44 \;cm}^{-1}$   \\
energy of the transition $X^2 \Sigma^+ \rightarrow B^2 \Sigma^+$ & 0.155 Ry\\
\hline 
\end{tabular}
%\end{center}
\caption{\label{tab:NaArdata}
Basic properties of the NaAr molecule as they are used for
fine tuning the model parameters.
}
\end{SCtable}
There are three data points and three parameters to be adjusted, namely
$A_{\rm el}$, $\beta_{\rm el}$, and $r_{\rm el}$. It turns out that these
three parameters compensate each other to some extent. One can, e.g.  vary the
potential height $A_{\rm el}$ and obtain about the same fit with re-tuning the
other two parameters. Fortunately, we find a reasonable fit to the wanted data
in spite of that restricted flexibility. Moreover, we exploit the freedom in
the choice of $A_{\rm el}$ to produce the softest reasonable core potential.
It is obvious that such a subtle adjustment depends crucially on the level of
approach. We thus obtain two different sets for quantum calculation and for
Vlasov LDA, similar as it was the case already for the Na ion-electron
pseudo-potential. The final parameters are summarized in table
\ref{tab:Arelparams}.
\begin{SCtable}[0.5]
%\begin{center}
\begin{tabular}{|l|ccc|}
\hline
  & $A_{\rm el}$ & $\beta_{\rm el}$ & $r_{\rm el}$\\
\hline
 quantum mechanical LDA & 0.47 & 1.6941\,/a$_0$ & 2.2 a$_0$ \\ 
 Vlasov-LDA & 0.4 & 1.44\, /a$_0$ & 1.8 a$_0$\\
\hline
\end{tabular}
%\end{center}
\caption{\label{tab:Arelparams}
Parameters for the ion-electron pseudo-potential in Na
according to the local form (\ref{eq:PsP}). Different sets are to be used for
quantum mechanical LDA and for semi-classical Vlasov-LDA.
}
\end{SCtable}

\section{Dynamical equations and their solution}
\label{sec:solve}

\subsection{The equations of motion}

The energy functional once specified (see section \ref{sec:enfundetail})
determines structure and dynamics of the coupled system.  This follows
standard techniques throughout. We thus give here only a very brief summary.
More details can be found in \cite{Rei03a,Rei03a} and specifically for
the semi-classical approach in \cite{Dom00a,Gig02}. 

The equations of motion are determined variationally. 
They read
\begin{subequations}
\begin{eqnarray}
  \imath\partial_t\varphi_n
  &=&
  \left(\frac{\hat{\bf p}^2}{2m_{\rm el}}
  +
  U_{\rm KS}\right)\varphi_n
  \quad,
\label{eq:KS}\\
  U_{\rm KS}({\bf r})
  &=&
  \frac{\delta E_{\rm total}}{\delta\rho_{\rm el}({\bf r})}
  \quad,
\end{eqnarray}
\begin{equation}
  \partial_t{\bf R}
  = 
  \nabla_{\bf P}E_{\rm total}
  \quad,\quad
  \partial_t{\bf P}
  = 
  -\nabla_{\bf R}E_{\rm total}
  \quad,\quad
  {\bf R}\in\left\{{\bf R}^{\mbox{}}_I,{\bf R}^{\mbox{}}_a,{\bf R}'_I\right\}
  \quad,
\label{eq:MD}
\end{equation}
for the case of quantum-mechanical propagation of electrons.
Only eq. (\ref{eq:KS}) is to be replaced by the Vlasov equation 
for the phase-space distribution $f({\bf r},{\bf p},t)$, i.e.
\begin{equation}
  \partial_t f
  +
  \left\{\frac{\hat{\bf p}^2}{2m_{\rm el}}+U_{\rm KS},f\right\}
  =
  0  
\label{eq:Vlasov}
\end{equation}
where $\{..,..\}$ is the classical Poisson bracket. Note that the
self-consistent Kohn-Sham potential $U_{\rm KS}$ using the LDA energy
functional is employed. That is why the scheme is called Vlasov-LDA.  In fact,
the above equations of motion describe electronic dynamics, in terms of TDLDA
or Vlasov-LDA, coupled to classical MD for ions and atoms. The whole scheme
is then TDLDA-MD in the quantum case or Vlasov-LDA-MD in the semi-classical
approximation.

It is the great advantage of the semi-classical approximation that it allows
to include dynamical electronic correlations through an \"Uhling-Uhlenbeck
collision term. This extends Vlasov-LDA to the Vlasov-\"Uhling-Uhlenbeck
scheme (VUU). The eq. (\ref{eq:Vlasov}) is then extended to
\begin{equation}
  \partial_t f 
  +
  \left\{\frac{\hat{\bf p}^2}{2m_{\rm el}}+U_{\rm KS},f\right\}
  =
  I_{\rm UU}
  \quad,
\end{equation}
\begin{eqnarray}
  I_{\rm UU}(\mathbf{r},\mathbf{p}_1,t) 
  &=&
  \int d^3p_2\,d^3p_3\,d^3p_4\,
  W(12,34) 
  \ (f^{\rm out}_{1,2}f^{\rm in}_{3,4}-f^{\rm in}_{1,2} f^{\rm out}_{3,4}) 
  \quad,
\label{eq:iuu}\\
  W(12,34) 
  &=&
  \frac{1}{m_{\rm el}^2} \ \frac{d\sigma}{d\omega} \
  \delta(\mathbf{p}_1+\mathbf{p}_2-\mathbf{p}_3-\mathbf{p}_4) \
  \delta(E_1+E_2-E_3-E_4) 
  \quad,
\\
  f^{\rm in}_{i,j} 
  &=&
  f(\mathbf{r},\mathbf{p}_i,t)f(\mathbf{r},\mathbf{p}_j,t)
  \quad,
\\ 
  f^{\rm out}_{k,l}
  &=& 
  (1- (2 \pi \hbar)^{3} \frac{f(\mathbf{r},\mathbf{p}_k,t)}{2}) \
  (1- (2 \pi \hbar)^{3} \frac{f(\mathbf{r},\mathbf{p}_l,t)}{2}) 
  \quad,
\label{eq:volelout}
\end{eqnarray}
\end{subequations}
where $i\equiv({\bf r},{\bf p})$ is used as an abbreviation and
$d\sigma/d\omega$ is the in-medium electron-electron cross section.  We use
here the cross section which had been evaluated for Na clusters
\cite{Dom00a,Gig02}. 

The numerical solution of the coupled equations of motion employs different
methods for the different ingredients \cite{Cal00}. The quantum-mechanical
Kohn-Sham equation (\ref{eq:KS}) is solved by the time-splitting method
\cite{Fei82}. The classical MD (\ref{eq:MD}) for ions and atoms is propagated
by the (velocity) Verlet algorithm \cite{Ver67}. The handling of the Vlasov
equation (\ref{eq:Vlasov}) and associated collision term
(\ref{eq:iuu}-\ref{eq:volelout}) is more involved. It will be detailed in the
next subsection \ref{sec:testpart}.

The static solution has to be provided before any dynamics can take off.
The ionic/atomic configuration is found by a mix of simulated annealing and
direct gradient towards the energy minimum.
Standard methods for the stationary Kohn-Sham equation apply in case of
the quantum mechanical treatment.
Most involved is again the semi-classical case. A stable ground state
distribution $f_0({\bf r},{\bf p})$ for the Vlasov equation is
determined by the Thomas-Fermi method \cite{Dom00a,Gig02}. It is to be
noted that the density and mean field entering Thomas-Fermi has to be
properly augmented with the test-particle folding (see next section)
in order to produce consistent input for the Vlasov solution with the
test-particle method.

\subsection{Test particle method}
\label{sec:testpart}

The phase-space distribution is a function in six-dimensional phase space.
Its representation on a (six-dimensional) grid is even today beyond the
capacity of computers. One thus uses a representation in terms of
test-particles. This technique had been imported for cluster physics from the
well developed nuclear physics case \cite{Ber88}. 
The smooth one-body phase-space distribution $f$ is represented by a swarm of
pseudo-particles each one having a certain width, i.e.
\begin{subequations}
\label{eq:testpansatz}
\begin{eqnarray}
  f(\mathbf{r},\mathbf{p},t) 
  &=&
  \frac{N_{\rm el}}{N_{\rm pp}}\sum^{N_{\rm pp}}_{i=1}
  g_{r}\big(\mathbf{r}-\mathbf{r}_i(t)\big) 
  g_{p}\big(\mathbf{p}-\mathbf{p}_i(t)\big) 
  \quad,
\\
  g_{x}(\mathbf{x}) 
  &=& 
  (2\pi\sigma_x^2)^{-3/2} \ \exp
  \big(-\frac{\mathbf{r}^2}{2 \sigma^2_x}\big)  
  \quad,\quad
  x\in\{r,p\}
  \quad,
\label{eq:gx}\\
  \sigma_r\sigma_p
  &=&
  \hbar
  \quad,
\label{eq:widrp}
\\
  \sigma_r
  &=&
  1.7836 \,{\rm a}_0
  \quad.
\label{eq:sigmar}
\end{eqnarray}
\end{subequations}
The folding widths $\sigma_r$ and $\sigma_p$ are free parameters of
the method. Relation (\ref{eq:widrp}) is connected to the Husimi
picture \cite{Hus40} which gives the elementary distribution $g_rg_p$
the interpretation of a minimal quantum-mechanical wavepacket being
used as means to derive and justify the semi-classical limit in the
smoothest way \cite{Dom97b,Pru78,Tak89}.  There remains $\sigma_r$ as
free parameter. It can be used to accommodate quantum-mechanical
binding properties through the Vlasov approximation as good as
possible and it has been used successfully that way for free
clusters in jellium approach \cite{Fer96,Pla99,Dom00a} as well as with
detailed ionic background \cite{Gig01a,Fen04,Gig02}.  We use the value
(\ref{eq:sigmar}) which provides a good reproduction of free neutral
Na clusters.

Inserting the ansatz (\ref{eq:testpansatz}) into the Vlasov equation
(\ref{eq:Vlasov}) yields the following equation of motion for the test
particles 
\begin{subequations}
\begin{eqnarray}
  \partial_t{\mathbf{r}}_i(t)
  &=&
  \mathbf{p}_i/m_{\rm el}
  \quad,
\\
  \partial_t{\mathbf{p}}_i(t)
  &=&
  -{\bf {\nabla}}_{r_i}U_{{\rm KS},g}
  \quad,
\label{eq:proptpp}
\\
  U_{{\rm KS},g}
  &=&
  g_r\star U_{\rm KS}
  =
  \int d^3r'  g_r(\mathbf{r}-\mathbf{r'})U_{\rm KS}({\bf r}')
  \quad.
\label{eq:potfold}
\end{eqnarray}
The Kohn-Sham potential is obtained from the ionic/atomic positions and
the actual electron density computed with eq. (\ref{eq:rhoelV}) which
reads in the test-particle representation
\begin{equation}
  \rho_{\rm el}({\bf r},t)
  =
  \frac{N_{\rm el}}{N_{\rm pp}}\sum^{N_{\rm pp}}_{i=1}
  g_{r}\big(\mathbf{r}-\mathbf{r}_i(t)\big) 
  \quad.  
\label{eq:denstp}
\end{equation}
\end{subequations}
Note that the $g_r$ folded Kohn-Sham potential (\ref{eq:potfold}) is employed
in the test-particle propagation (\ref{eq:proptpp}).  This complies with the
Husimi picture and it allows an efficient handling of density and mean-field
on a spatial coordinate-space grid. The density is accumulated on the grid
with eq. (\ref{eq:denstp}) using the $g_r$ so to say as interpolation
functions. The Coulomb and exchange-correlation potentials are determined on
the grid composing the Kohn-Sham potential together with the potentials from
ions and atoms. The forces on the test-particle are retrieved from the grid
again with the help of the folding functions $g_r$. This is the practical side
of eq. (\ref{eq:potfold}).

The finite representation in terms of test-particles induces a principle
long-time instability towards a final Boltzmann equilibrium distribution.
This is unphysical because it violates the Pauli principle
\cite{Rei95a,Rei95e}. The critical time depends on folding width and number of
test particles. The folding width is fixed by the adjustment to quantum
features. It remains the particle number. We chose $N_{tp}=4000$ per electron.
This provides sufficient stability over the time
scale considered (200 fs). The collision term (\ref{eq:iuu}) in VUU contains
the Pauli blocking. This provides long time stability of the propagation.

The practical evaluation of the collision term is already simplified by the
test particle. The nine-fold momentum-space integration is replaced by
four-fold summation over test-particles, each factor $f_i$ contributing one
sum. The scattering process connects two incoming momenta with two outgoing
ones, $({\bf p}_1,{\bf p}_2)\longrightarrow({\bf p}_3,{\bf p}_4)$. The
outgoing momenta are fixed by momentum and energy
conservation laws up to a scattering angle.
The scattering angle is chosen from the unit sphere at random.

\subsection{Excitation mechanism and observables}
\label{sec:excobs}

Typical cluster excitations are promoted by laser pulses or collision with
highly charged ions. Both can be described as external time-dependent Coulomb
fields acting on the cluster. The coupling to the field $V_{\rm ext}({\bf
r},t)$ is provided for that purpose in the basic energy functional
(\ref{eq:etotal}). A laser pulse, e.g., is characterized by 
$V_{\rm ext}\propto\hat{D}\cos{(\omega t)}f_{\rm profile}(t)$. 
In this first survey, we are interested mainly on general spectral
properties. These are explored in the limiting case of an arbitrarily short
laser pulse. It is realized by an instantaneous initial boost of the
whole electron cloud which reads, in the Vlasov case,
\begin{equation}
  f({\bf r},{\bf p},t\!=\!0)
  =
  f_0({\bf r},{\bf p}+{\bf b})
\end{equation}
where $f_0$ is the ground state distribution and ${\bf b}$ is the boost
momentum. In the quantum case the boost is applied to each electron 
orbital as 
\begin{equation}
  \varphi_\alpha({\bf r})
  \longrightarrow
  \varphi_\alpha({\bf r},t\!=\!0)
  =
  e^{\imath b{\bf r}}\varphi_\alpha({\bf r}) \qquad .
\label{eq:boost}
\end{equation}
The size of ${\bf b}$ determines the dynamical regime. Small 
${\bf b}$ explore the regime of linear response.

The dipole spectrum (optical absorption) is obtained from the subsequent
dipole signal ${\bf D}(t)=\int d^3r$ $\rho_{\rm el}({\bf r},t)$ $e{\bf r}$.  It is
recorded over the whole time evolution. The dipole strength is finally
obtained from the imaginary part of the Fourier transform, $\Im\{\tilde{\bf
D}(\omega)\}$. The three vector components correspond to the modes in $x$-,
$y$-, and $z$-direction. For details see \cite{Rei03a,Cal00,Cal97b}.

A further key observable is electron emission. The computation of the number
of emitted electrons $N_{\rm esc}$ is straightforward. In TDLDA, we use
absorbing boundary conditions \cite{Cal00}. The lost electrons are determined
as the complement of the still remaining electrons $N_{\rm esc}=N(0)-\int
d^3r\,\rho_{\rm el}({\bf r},t)$.  In Vlasov-LDA, we draw a large sphere around
the simulation area with radius $20 \,{\rm a}_0$ and count explicitely
all test-particles leaving that sphere.

\section{Benchmark NaAr}
\label{sec:benchNaAr}

Most of the parameters entering our model are taken from other
grounds, i.e. from values published elsewhere and adjusted to
independent data. The various constituents contribute strongly
counteracting forces. This calls for some freedom to final
fine-tuning. We exploit the anyway not so easily determined
electron-Ar core potential for that purpose. The form (\ref{eq:VArel})
is taken from quantum-chemical investigations \cite{Dup96}. The
parameters are given free for fine-tuning. Thereby we consider the
Na-Ar dimer as benchmark. The fully quantum mechanical TDLDA includes
the Van-der-Waals interaction (\ref{eq:EvdW}) while the semi-classical
approach does not have that term. The missing contribution is then
mimicked to some extent by the readjustment of the free parameters of
the electron-Ar potential.
\begin{SCfigure}[0.5]
{\epsfig{figure=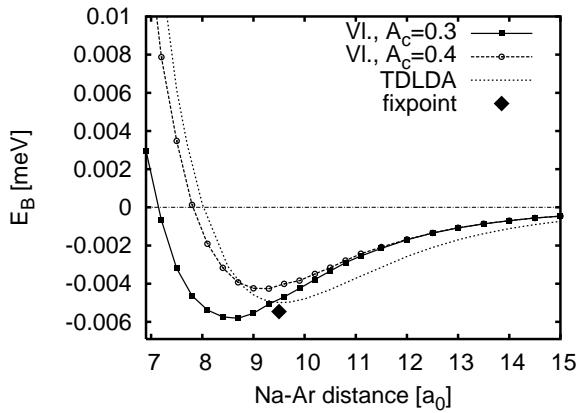,width=8cm}}
\caption{\label{fig:Na_Ar_pes}
The Born-Oppenheimer surface for the NaAr dimer computed semi-classically with
the test-particle representation and compared with
full TDLDA. The parameters for both TDLDA and semi classics
are given in table \ref{tab:Arelparams}.
In the semi-classical case, two different core heights in the Ar-electron core
potential (\ref{eq:VArel}) are considered. The value $A_c=0.4$ is the
final standard.
The filled rhombus indicates the fitting point as listed in table
\ref{tab:NaArdata}. 
}
\end{SCfigure}
Figure \ref{fig:Na_Ar_pes} shows the Born-Oppenheimer surfaces for the Na-Ar
molecule. The TDLDA fit complies nicely with the data point. The
semi-classical results manage to reach that approximately.  The results from
two different core heights serve to demonstrate the variances one can explore
in the adjustment: while reducing the core height, one can shift the minimum
closer to the wanted value, but at the same time one finds the bonding energy
moving off the optimum. Whatever one does, one finds that the semi-classical
Born-Oppenheimer surfaces fall off much more quickly than the TDLDA one.  Note
that the asymptotic in case of TDLDA is dominated by the Van-der-Waals
attraction $\propto r^{-6}$. Vlasov-LDA does not have that term and thus is
bound to display a different asymptotics. Altogether, one sees that it is by
no means trivial that a good adjustment can be found at all. Thus the TDLDA
fit is a remarkable agreement and we are also satisfied with the
semi-classical results which manage to find a fair agreement for the binding
point. One has to keep in mind that the stronghold of Vlasov-LDA and VUU are
the energetic processes. For these it suffices to provide a stable starting
ground-state configuration which reproduces roughly realistic binding
properties. This is achieved by the choice of parameters as listed in table
\ref{tab:Arelparams}. Thereby we have preferred the larger core height which
provides the correct bond length while giving a slightly too small
binding energy. Note that the semi-classical adjustment was done without
explicit Van-der-Waals energy. This term is derived from quantum
mechanical perturbation theory. Its implementation in a semi-classical
environment requires deeper theoretical development and what we have
found  would increase the computational cost by an order of
magnitude. That is not very useful for an approach which concentrates
on processes far above the low energy scale of binding.

A further benchmark point for TDLDA is the energy for the $X^2 \Sigma^+
\rightarrow B^2 \Sigma^+$ transition at the bonding configuration. The goal is
a transition energy of 0.155 Ry (see table \ref{tab:Arelparams}); the TDLDA
result with the optimized parameters is 0.164 Ry, a satisfying agreement.
This benchmark cannot be applied for the
semi-classical approach because there are no quantized single-particle states
and thus no specific transitional energy spectra. We will apply Vlasov-LDA to
metal clusters. These have the Mie surface plasmon as prominent excitation
mode and this mode is determined by collective flow rather than by
single-particle excitations. Thus we can expect a good reproduction of the
Mie plasmon in spite of the missing single-particle adjustment. This is indeed
the case as many previous applications for free metal clusters have
exemplified \cite{Fer96,Pla99,Dom00a,Gig01a,Gig02,Fen04}. We have yet to see
how the model performs for embedded clusters.

\section{The Na$_8$\@Ar$_N$ system}
\label{sec:Na8}

\subsection{Structure}
\label{sec:struc}

The ground state configurations involve 8--220 ions and atoms. This is
hard to visualize in all detail. The basic structure, however, is
simple \cite{xx1,xx2}. There exist three close isomers for the Na$_8$
clusters \cite{xx2}.  We consider here the Na$_8$ cluster in the shape
consisting out of two rings with each four electrons. The Ar matrix
was originally arranged in radial shells. It is slightly perturbed by
the embedded cluster. But the shells still remain a useful guide and
sorting scheme. Thus we will characterize the Na$_8$ cluster by its
global root-mean-square radius and the Ar matrix is visualized as
distribution of atoms in radial shells.

\begin{SCfigure}[0.5]
{\epsfig{figure=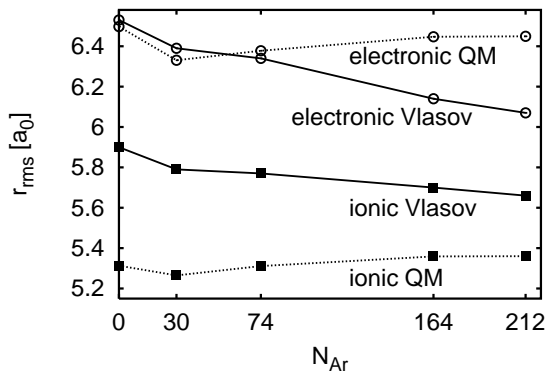,width=7.5cm}}
\caption{\label{fig:Na8_geom}
Ionic and electronic radii of a Na$_8$ cluster embedded in
Ar matrices of different size. 
Results from Vlasov and from quantum mechanical
treatment are compared. 
}
\end{SCfigure}
Figure \ref{fig:Na8_geom} shows the radii for the (embedded)
Na$_8$. The ionic radii are generally smaller than the electronic ones because
the electronic cloud reaches farther out than the well localized ions. The
difference between electronic and ionic radius is caused mainly by the
Gaussian folding with width $\sigma_r$. It is thus nearly constant.  It is of
the same order (although somewhat smaller) as the difference for the quantum
mechanical case. The latter shows an interesting detail: it is practically
constant for cases with Ar matrix but jumps by 0.2 a$_0$ for the free cluster.
This is due to the onset of core repulsion, similar as the first step for
Vlasov LDA. In the further evolution,
the radii show a different trend. While they slowly grow with $N_{\rm
Ar}$ in the quantum mechanical case, they shrink for Vlasov-LDA. This hints at
a different mix of interactions with the Ar atoms. The Ar core repulsion is
felt more strongly for the test-particle in Vlasov-LDA because these are
associated rigidly with a Gaussian distribution. The repulsive core wants to
push away the tail and thus pushes the whole Gaussian. This works much
different for the quantum mechanical case. The electron cloud can react more
``elastically''. It allows for a dip near the Ar core and evolves else-wise
almost unperturbed around it. Thus the dipole attraction can act more freely
to pull the electronic tail outward.
\begin{figure}
\centerline{\epsfig{figure=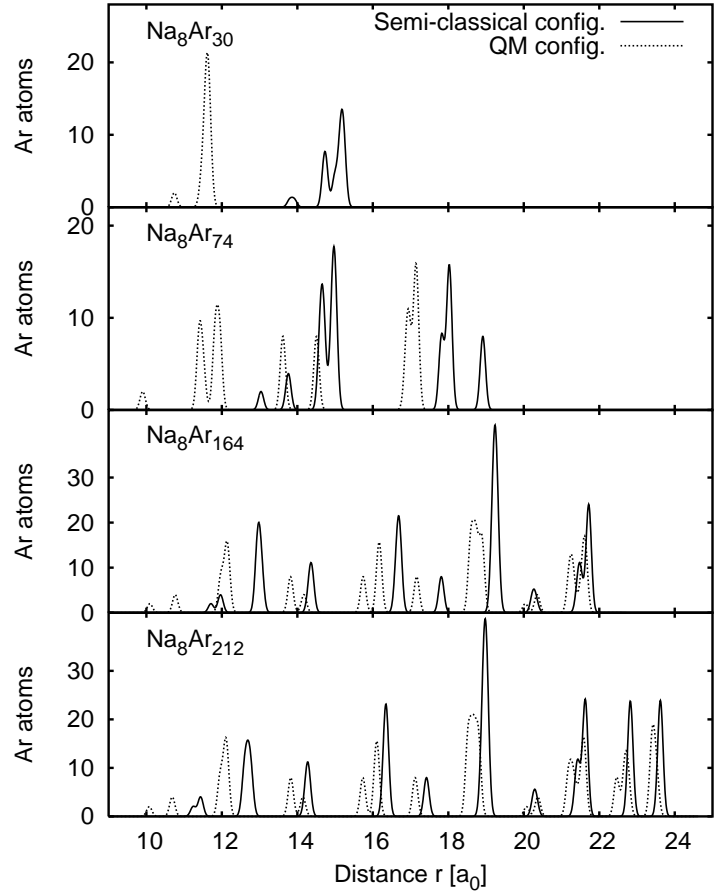,width=10cm}}
\caption{\label{fig:Na8_Arshells}
The distribution of radial shell in Ar matrices of different sizes
for Vlasov-LDA and for a fully quantum mechanical treatment.
}
\end{figure}
Figure \ref{fig:Na8_Arshells} shows the Ar part of the combined systems in
terms of the radial atom distribution. The differences between quantum LDA and
Vlasov-LDA structures are dramatic for the small Ar matrices. They converge to
each other with increasing system size, similarly to the convergence 
observed in the case of Na properties (Figure \ref{fig:Na8_geom}).
The difference is again due to the
rigidity of the fixed Gaussian electron distribution for Vlasov-LDA. This
exaggerates core repulsion and diminishes dipole attraction. Moreover, we
are missing the long range Van-der-Waals attraction (see figure
\ref{fig:Na_Ar_pes}). The pressure from the outside shells and the increasing
dipole attraction from the far shells produce then the compression towards the
quantum mechanical (and thus more realistic) structures.
The example demonstrates the limitations of the test-particle method. The
smooth Gaussian folding implied in the test-particle representation was
perfectly suited for the soft ionic pseudo-potentials of a free metal cluster
(and the more so for the jellium approach to a free cluster). But the inherent
rigidity of the electronic test-particle distribution becomes a hindrance in
connection with steep potentials as they appear here in terms of the repulsive
Ar cores. The Vlasov-LDA is certainly not appropriate for detailed studies of
subtle binding effects. We find, however, that the structures for larger
matrices come out sufficiently well to provide a laboratory for dynamical
studies in the non-linear regime. Remember that 
VUU versus Vlasov-LDA is presently the 
only way to estimate microscopically the effect of dynamical electron
correlations on the resonance excitations.

\subsection{Dynamics}

\begin{SCtable}[0.5]
%\begin{center}
\begin{tabular}{|l|l|}\hline
domain & excitation energy\\
\hline
linear & 0.3 eV \\
semi-linear & 4.1 eV\\
non-linear &  13.5 eV\\
\hline
\end{tabular} 
\caption{\label{tab:domenerg}
Excitation energies for different domains.}
%\end{center}
\end{SCtable}
In order to explore the basic dynamical properties, we employ the most
simple excitation mechanism of an instantaneous boost, as explained in
section~\ref{sec:excobs}. It has only one parameter, the boost
strength, which allows to scan the various dynamical regimes. We pick
a typical strength for each regime as listed in table
\ref{tab:domenerg} and characterized by the initial excitation
energy. To put that in perspective, one may compare it with the
typical energies in the system: at the side of the Na cluster we have
the ionization potential with IP$\approx 3.5-4.5$ eV and the energy of
the Mie surface plasmon with $\omega^{\rm(Mie)}\approx 2.6-2.9$ eV.
At the side of the Ar atoms, we have their first electronic
excitation and the IP above 20 eV. The linear regime is far below
anyone of these scales, the semi-linear just of the same order, and
the non-linear regime safely above (multi-plasmon, strong emission).

\begin{SCfigure}[0.5]
\epsfig{figure=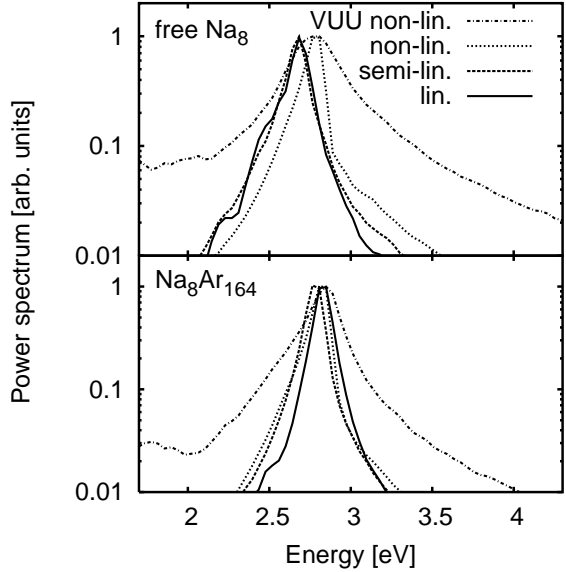,width=8cm}
\caption{\label{fig:spec_detail}
Results of Vlasov calculations for the
power spectrum of the dipole mode along $z$ axis for Na$_8$ (upper
panel) and for
Na$_8$@Ar$_{164}$ (lower panel) at different excitation energies.
The spectra are normalized to peak maximum at 1 for better comparison. 
The dipole moment has been recorded for 100 fs.
}
\end{SCfigure} 
The spectral properties from Vlasov-LDA (and VUU) are shown in figure
\ref{fig:spec_detail}
in terms of the power spectrum of the dipole mode along
$z$-axis (which corresponds to the 
approximate symmetry axis of the Na$_8$ cluster). The general
pattern are much similar to previous studies on Na$_9^+$
\cite{Gig01b,Gig02}. There is a clear
resonance peak under any conditions. It is not destroyed by high excitations.
These well developed resonance features persist very well also for the
embedded cluster. One may even find the pattern somewhat cleaner than for the
free cluster. At second glance, one sees trends and developments. There is,
e.g., the much larger width in case of VUU which is obviously generated by the
electron-electron collisions. We will now work out the trends in terms of
the key features of these simple spectra: the peak position
$\omega^{\rm(Mie)}$ and the width as full width at half maximum (FWHM).

\begin{SCfigure}[0.5]
\epsfig{figure=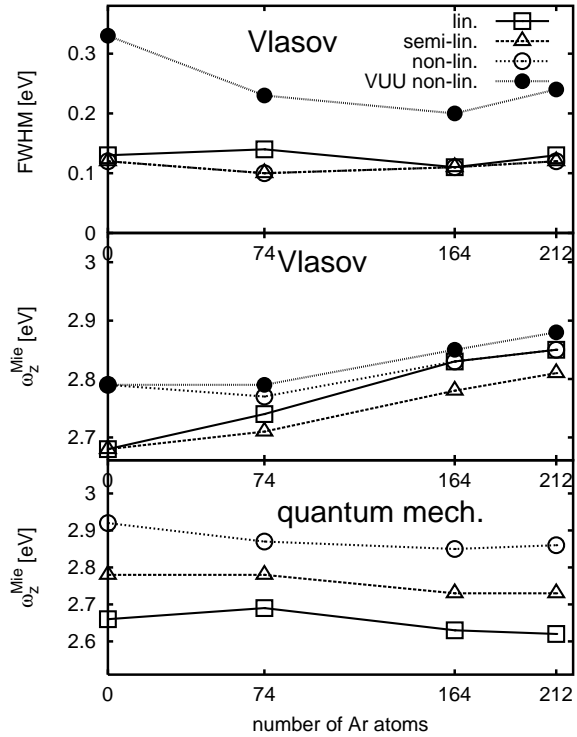,width=8cm}
\caption{\label{fig:spec_trends}
Position of the Mie plasmon (lower two panels) and its full width at half
maximum (FWHM, uppermost panel) for Na$_8$\@Ar$_{n}$ at various excitations
as indicated.
Lowest panel: quantum mechanical results. Upper two panels: results from
Vlasov-LDA (or VUU, respectively).
}
\end{SCfigure} 
Figure \ref{fig:spec_trends} shows these basic features in case of the
system Na$_8$\@Ar$_{n}$ ($n < 212$) for quantum 
mechanical calculations, Vlasov-LDA, and VUU.
The peak frequencies (two lower panels) depend only very weakly on the Ar
matrix in all cases.  The remaining small trend shows a red shift with
increasing matrix size in the quantum mechanical results. That is due to the
fact that the additional Ar shells enhance the attractive dipole
interaction. The Vlasov results indicate the opposite trend. That is due to
the decreasing radius of the inner Ar shell (see figure
\ref{fig:Na8_Arshells}) which gives more weight to the increasing core
repulsion. 
The changes with increasing excitation energy differ. The quantum mechanical
results show a sizeable and systematically increasing blue-shift. It can be
explained by the increasing ionization which, in turn, deepens the binding
potential and thus increases the restoring force for the dipole oscillations
\cite{Cal97b}. The same trend is hinted in the Vlasov results for the free
Na$_8$ shown in the upper panel of figure \ref{fig:spec_detail}. But the
Vlasov results for the embedded cluster (middle panel of figure
\ref{fig:spec_trends}) behave differently in that the step from the linear to
the semi-linear regime shows a tiny (or no) red shift instead of a blue
shift. The reasons for that are not yet fully clear. The step into
the non-linear regime then produces the blue-shift. Finally, the step to VUU
has negligible effect on the peak position as one would have expected.

The uppermost panel of figure \ref{fig:spec_trends} shows the widths of the
resonance peaks in the Vlasov cases. The spectral analysis uses a recording
time of 100 fs and employs a filtering to avoid artefacts from insufficiently
relaxed signals \cite{Pre92}. This sets the energy resolution of the analysis
to 0.1 eV. The width of the Vlasov results is hidden in the resolution except
for the non-linear case which shows already some enhancement due to a
dynamical spreading (through ongoing ionization) of the spectra. However, the
collision term in VUU adds a substantial bit of 0.2 eV for free Na$_8$ and 0.1
eV for the embedded clusters. It is a bit surprising to see a reduction of the
width when the matrix is around. This hints at a somewhat more harmonic
potential for the embedded cluster. The effect deserves a deeper analysis
which, however, goes beyond scope and limits of this paper.

\begin{SCfigure}[0.5]
\epsfig{figure=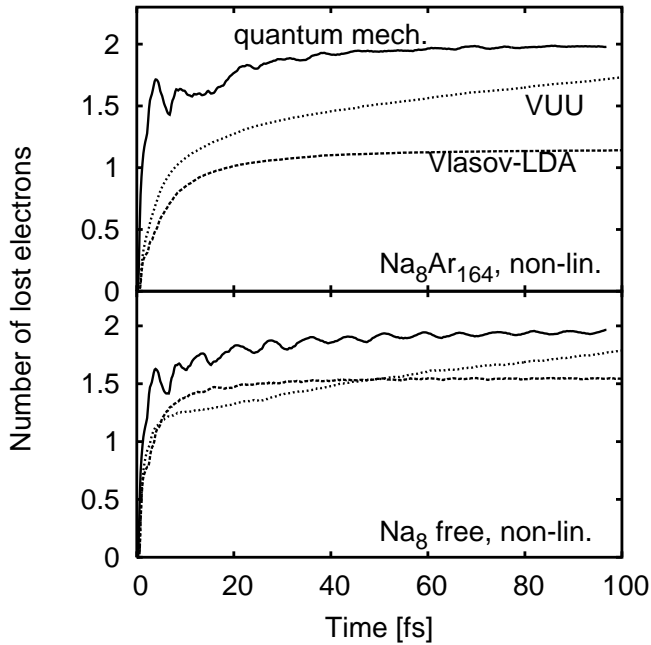,width=9cm}
\caption{\label{fig:nesc_detail}
Time evolution of the number of emitted electrons $N_{\rm esc}$ 
after strong initial boost with excitation energy 13.5 eV (non-linear
regime). Results from quantum mechanical TDLDA, Vlasov-LDA, and VUU
are compared. Upper panel: for the system Na$_8$@Ar$_{164}$.
Lower panel: for free Na$_8$.
}
\end{SCfigure} 
Figure \ref{fig:nesc_detail} shows the time evolution of ionization (number of
emitted electrons) in case of the non-linear excitation.  The pattern for
quantum-mechanical TDLDA and Vlasov-LDA are similar: there is a steep initial
increase from direct electron emission and then a quick bending over to an
almost flat trend. The results agree even quantitatively for the free cluster.
For the embedded system, on the other hand, the agreement may be just
acceptable but is certainly less perfect with Vlasov-LDA showing about half of
emission from TDLDA. The reason is that cores of the surrounding Ar atoms
build up a narrow potential barrier of about 1 eV height. This barrier
reflects in Vlasov-LDA all test particles which do not have sufficient kinetic
energy while quantum mechanical electrons in TDLDA can easily tunnel
through. More violent excitations produce a larger fraction of faster
electrons and will thus diminish that difference. This confirms once again
the rule that semi-classical approaches become increasingly valid with
increasing excitation energy.
The VUU results in figure \ref{fig:nesc_detail} show a qualitatively different
trend. There remains a slope which hints at significant delayed emission of
thermalized electrons. The free Na$_8$ shows the well understood deal: the VUU
curve remains below Vlasov-LDA at early times because direct emission is
suppressed in favor of internal thermalization while it crosses the Vlasov
line at later times when thermal emission takes over \cite{Dom00a}.  In case
of the embedded cluster, one finds the VUU emission enhanced in all time
domains. We see here again a classical effect. 
The part of thermal emission in VUU is
isotropically distributed. The electrons have thus more chances to escape
through the saddles between the cores. Moreover, the more effective
thermalization in VUU allows to fill more systematically the Boltzmann tail of
the kinetic energy distribution until some electrons have gathered sufficient
energy to surmount the extra barrier.

\begin{SCfigure}[0.5]
\epsfig{figure=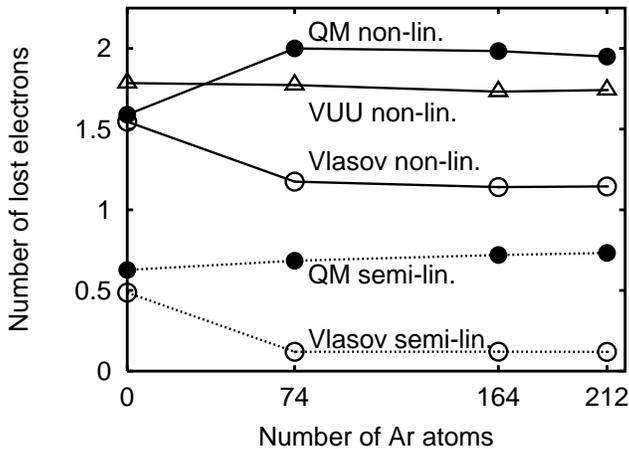,width=9cm}
\caption{\label{fig:nesc_trends}
Final number of emitted electrons $N_{\rm esc}$ 
after initial boost in semi-linear and non-linear regime
(see table \ref{tab:domenerg}).
Results from quantum mechanical TDLDA, Vlasov-LDA, and VUU
are compared.
}
\end{SCfigure} 
Figure \ref{fig:nesc_trends} shows a summary of asymptotic values for
ionization for different regimes,
methods and system sizes. There is little dependence on the size of the Ar
matrix, except for the too small $N_{\rm Ar}=30$.  Thus we recognize many
features from the previous figure.  We see again the strong suppression of
(direct) emission in Vlasov-LDA. It is even more dramatic in semi linear
regime (factor 8 suppression!) because it has a larger fraction of slow
electrons which cannot surmount the first Ar barrier.  We also find again a 
systematic trend of enhanced emission in VUU due to its larger contribution
from isotropic emission and from the Boltzmann tail.
Altogether, the results for electron emission hint that the propagation of
slow electrons in the Ar matrix is much different between quantum TDLDA and
semi-classical approaches. The latter still provide useful guidelines and
presently the only chance to study effects of electron-electron collisions. One
should, however, take care to remain in the regime of strong excitations where
slow electrons are less important.

\section{Conclusion}

In this paper, we have presented a model for the description of the
dynamics of an embedded metal cluster (Na) in a rare gas matrix (Ar).
The model employs a microscopic description of three basic
ingredients: cluster valence electrons, cluster ionic cores, and
raregas atoms.  The cluster electrons are treated within
time-dependent density functional theory (DFT). The cluster ions are
treated classically. The raregas atoms are also treated as classical
particles together with their dynamical polarizability as intrinsic
degree of freedom. This model allows to handle a large variety of
dynamical situations at the side of the cluster, including electron
emission when non linear excitations are considered. The cluster
electrons can be treated at 2 levels of approximation within the realm
of DFT: fully quantum mechanical time-dependent local-density
approximation (TDLDA) and (semi-classical) Vlasov-LDA. It turns out
that both approaches give results which are qualitatively, sometimes
even quantitatively, comparable.  This holds in the stationary as well
as in the dynamical regime.  Fully fledged TDLDA reproduces quite well
experimental data and previous quantum chemistry approaches.  The
semi-classical approach also provides an acceptable description for
larger raregas matrices, which is to some extend a surprise in view of
the subtle energetic balance in these systems.

The semi-classical approach becomes particularly suited when
non-linear excitations are considered. It presently provides the
simplest and practically only way to deal with dynamical correlations
of the cluster electrons.  This is achieved by complementing the
original Vlasov equation by a Boltzmann-\"Uhling-Uhlenbeck collision
term yielding the Vlasov-\"Uhling-Uhlenbeck (VUU) model.  The
collisional correlations induce significant differences between the
Vlasov and VUU approaches when a sufficiently large energy is
deposited in the system, for free clusters as well as for embedded
ones.
The test cases considered make it desirable to account for these
dynamical correlations also in a quantum mechanical context.  This is,
however, an extremely difficult task still under development. In the
mean time the VUU approach has proven to provide a valuable tool
for investigations of the non linear dynamics of simple metal clusters
embedded in a rare gas matrix. Further explorations of the various
competing phenomena are underway.

\bigskip

\noindent
Acknowledgments: 
This work was supported by the DFG, project nr. RE 322/10-1,
the french-german exchange program
PROCOPE number 04670PG, the CNRS Programme "Mat\'eriaux" 
  (CPR-ISMIR), Institut Universitaire de France, 
  Humbodlt foundation and  Gay-Lussac price.

\bibliographystyle{unsrt} 
\bibliography{cluster,add2}

\end{document}